\documentclass[twocolumn,amssymb,10pt,unsortedaddress]{revtex4-1}
\usepackage[english]{babel}
\newcommand{\degree}{\ensuremath{^\circ}}
\newcommand{\iu}{{i\mkern1mu}}
\usepackage{amssymb}
\usepackage[dvips]{graphicx}
\usepackage{booktabs}
\usepackage[table]{xcolor}
\usepackage{tabularx}
\usepackage{colortbl}
\usepackage{graphicx}

\usepackage{color}
\definecolor{darkblue}{rgb}{0,0,0.5}
\definecolor{lila}{rgb}{0.3,0,0.3}
\definecolor{turq}{rgb}{0,0.1,0.4}
\definecolor{lightblue}{rgb}{0.7,0.7,0.9}
\usepackage{url} % that hyperlinks may be hyphenated correctly
\usepackage[pdftex,
 colorlinks=true,
 backref=page,
 linkcolor=darkblue, % usual links
 filecolor=red,
 citecolor=turq, % for bibliographic
 urlcolor=lila, % for Emails etc
 pdftitle={Delay and polarization routing of single photons},
 pdfauthor={Julian Maisch, Hueseyin Vural, Michael Jetter, Peter Michler, Ilja Gerhardt, Simone Luca Portalupi},
 pdfsubject={},
 pdfkeywords={},
 pdfpagelabels=true, % damit die Nummerierung stimmt
 breaklinks=false,
 plainpages=false,
 backref=false,
 bookmarks,
 bookmarksnumbered=true]{hyperref}

\begin{document}

\title{\Large Delay and polarization routing of single photons}

%\author{Julian Maisch$^{1}$, H\"useyin Vural$^{1}$, Michael Jetter$^{1}$, Peter Michler$^{1}$, Ilja Gerhardt$^{2}$ and Simone Luca Portalupi$^{1}$}
%\affiliation{$^{1}$Institut f\"ur Halbleiteroptik und Funktionelle Grenzfl\"achen (IHFG), Center for Integrated Quantum Science and Technology (IQ$^{ST}$) and SCoPE, University of Stuttgart, Allmandring 3, D-70569 Stuttgart, Germany}
%\affiliation{$^{2}$3. Institute of Physics, Center for Integrated Quantum Science and Technology (IQ$^{ST}$), University of Stuttgart, Pfaffenwaldring 57, D-70569 Stuttgart, Germany}

\author{Julian Maisch$^{1}$} 
\email[]{j.maisch@ihfg.uni-stuttgart.de}

\author{H\"useyin Vural$^{1}$} 

\author{Michael Jetter$^{1}$} 

\author{Peter Michler$^{1}$} 

\author{Ilja Gerhardt$^{2}$}

\author{Simone Luca Portalupi$^{1}$}

\affiliation{$^{1}$Institut f\"ur Halbleiteroptik und Funktionelle Grenzfl\"achen (IHFG), Center for Integrated Quantum Science and Technology (IQ$^{ST}$) and SCoPE, University of Stuttgart, Allmandring 3, D-70569 Stuttgart, Germany}
\affiliation{$^{2}$3. Institute of Physics, Center for Integrated Quantum Science and Technology (IQ$^{ST}$), University of Stuttgart, Pfaffenwaldring 57, D-70569 Stuttgart, Germany}

% \ead{i.gerhardt@fkf.mpg.de}

\begin{abstract}
The full control of single photons is important in quantum information and quantum networking. A convenient storage device for photons is the key to memory assisted quantum communication and computing. While even a simple optical fiber can act as a convenient and reliable storage device, its storage time is tightly fixed and cannot be adapted. Therefore, the photon storage should ideally be actively controllable by external means, such as magnetic or electric control fields. In order to multiplex several photons, an active routing would also be desirable. Here we show that single photons of a semiconductor quantum dot can be deliberately delayed by an atomic vapor. Also, the output path can be selected, depending on an external magnetic field. By selecting the input polarization of the photons and by aligning the external magnetic field of the hot atomic vapor, the delay-based storage can be fine tuned to a deliberate value. With an overall delay of 25~ns, we are able to fine tune by more than 600~ps. Depending on the input polarization, the photons are routed into different output ports. The experimental data is fully resembled by a theoretical model, which describes the group velocity delay under consideration of spectral diffusion and considers the complex refractive index of the atomic vapor. The present results enable the use of an atomic vapor as a wavelength selective delay and allows for routing the single photons according to their polarization and an external magnetic field.
\end{abstract}
\maketitle

\section{Introduction}
%% Intro on single photons and QIP
Single photons are an essential ingredient of modern quantum information processing. Encoding information into single photons will result in highly secure data transfer which becomes particularly appealing for the implementation of quantum cryptography and communication~\cite{townsend_el_1993}. To be beneficial in the up-scaling of the experimental complexity, their generation is ideally realized in a deterministic or turnstile way, such that a single photon can be generated on demand. This forms the key advantage of single emitter based single photon sources against so-called parametric down-conversion sources. Typically on-demand single photon sources originate from single atoms~\cite{kimble_prl_1977} and ions, over molecules~\cite{basche_prl_1992} to defect centers~\cite{gruber_s_1997} and quantum dots~\cite{michler_s_2000}. Since quantum dots (QDs) are based on semiconductor technology, they hold the promise to be integrated onto chip-scale devices. Furthermore their option to generate polarization entangled photons~\cite{akopian_prl_2006,young_njp_2006} opens the potential for a variety of quantum information schemes, which are enabled with this quantum phenomenon.

%% Hybrid systems
Current state-of-the-art QDs exhibit high brightness, high indistinguishability~\cite{unsleber_oe_2015,somaschi_np_2016,ding_prl_2016} and entanglement fidelity~\cite{michler_book_2017}. Quantum dots therefore represent a very appealing source of non-classical light. On the other hand, in many quantum applications, such as quantum repeaters, the implementation of a deterministic quantum memory is beneficial. At present, the relatively short coherence times of the QD's spin may limit their performance as storage media. In contrast, atomic systems with their high coherence and long storage times could provide this option~\cite{phillips_prl_2001}. Still, the implementation of a full storage experiment is a challenging task. For instance, the intensity mismatch between single photons and the required high power control fields complicates the experimental realization. Another limiting factor is the bandwidth mismatch, when e.g.\ broad photons have to be stored in a spectrally narrow medium. An intermediate step towards the on-demand realization is the implementation of an optical delay line within an atomic medium. First results with laser pulses~\cite{carruthers_joap_1969,grischkowsky_pra_1973,siddons_np_2009} prove that an atomic vapor has a large application potential~\cite{siddons_pra_2010}. Such experiments have set the basis for recent experiments where single photons were slowed down within alkali vapors~\cite{akopian_np_2011,siyushev_n_2014,wildmann_prb_2015,kiefer_apb_2016,vural_o_2018}. These experiments show impressingly the compatibility of single photons from quantum dots with their atomic counterparts. Still, photonic routing, or photon multiplexing~\cite{kaneda_a_2018} was not shown in these papers.

%% what is shown
Here we report on the experimental combination of implementing a fine-tunable delay and simultaneous photonic routing in a hot atomic cesium vapor. The experiments are conducted with a semiconductor quantum dot as an on-demand source of single photons. The delay and the routing is implemented by a 250~mm long Cs-vapor cell in a externally controllable magnetic field. Depending on the external magnetic field, the two polarization components of the light are affected differently. This feature can be deliberately controlled and fine-adjusted by the external magnetic field.

%%%%%%%%%%%%%%%%%%%%%%%%%%%%%%%%%%%%%%%%%%%%%%%
%% Theory part 
\section{Theory}
The refractive index in a hot atomic vapor is tied to the strong absorption lines and the dispersion in the medium. This well-know phenomenon is usually represented by a division into a real and an imaginary part of the refractive index -- the Kramers-Kronig relation. The index can be divided as $n=n'+\iu n''$. With the electric susceptibility, $\chi$, this is approximately equivalent to $n=1+2\pi \chi$, where $\chi$ is represented as

\begin{equation}
\chi=\frac{N e^2/2m\omega_0}{(\omega_0-\omega)-\iu \gamma}\, .
\end{equation}

\noindent
Here, $N$ represents the number of involved atoms, $e$ is the electron charge, and $\omega_0$ represents the transition and $\omega$ the laser frequency. $\gamma$ represents the radiative lifetime of the excited state. Therefore, in other words, the refractive index is represented by the two components
\begin{eqnarray}
n'=1+\frac{\pi N e^2}{2 m \omega_0 \gamma} \frac{2(\omega_0-\omega)\gamma}{(\omega_o-\omega)^2+\gamma^2}\\
n''=\frac{\pi N e^2}{2 m \omega_0 \gamma} \frac{\gamma^2}{(\omega_0-\omega)+\gamma^2} \, .
\end{eqnarray}

\begin{figure*}[htb]
  \includegraphics[width=\textwidth]{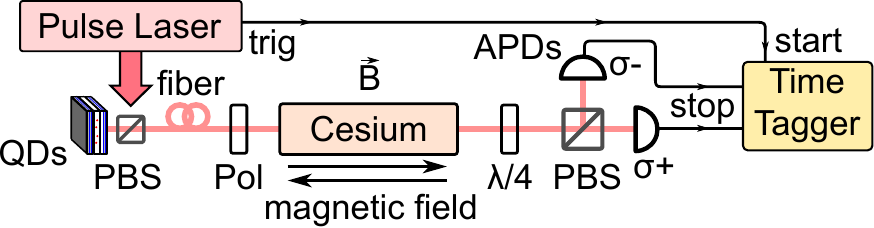}
  \caption{Experimental setup for time-correlated single photon counting (TCSPC). The quantum dot (QD) is resonantly excited. Through the polarizer (Pol) the single photons enter the cesium vapor cell linearly polarized. A variable magnetic field can be applied parallel or anti-parallel to the propagation direction. Behind the vapor cell the quarter wave plate ($\lambda /4$) projects the circular polarizations onto orthogonal linear components. These are separated onto the two APDs via the polarizing beamsplitter (PBS). The signals are recorded by time tagging.}
  \label{fig:fig01}
\end{figure*}

\noindent
%% The delay of photons
The group velocity inside a medium of group index is known to be given as $n_g=n'+\omega dn'/d\omega$. When the group index is larger than unity ``slow light'' can be observed~\cite{akopian_np_2011,siyushev_n_2014,vural_o_2018}, while for an index below unity ``fast light'' is expected~\cite{chu_prl_1982,kroh_a_2019}. 

%% magnetic fields
The Zeeman-effect is responsible for a spectral split of the atomic lines under the influence of a longitudinal magnetic field. This naturally influences not only the absorption but also the dispersive components of the refractive index. Both of the two split dispersion components act differently on circular polarized light. This holds consequently also for linearly polarized light, which represents as a linear-combination of two circular fields. In summary, this leads to an effective rotation of light which is based on the Faraday effect~\cite{gerhardt_ol_2018}. This has been used in Faraday filters and the symmetry breaking allows to utilize this effect for optical isolation on or close to the atomic resonance~\cite{weller_ol_2012}.

%% locking
Both schemes analyze the input of linear polarized light and rely on the effective rotation of the linear component by the atomic medium. A linear analyzer behind a vapor cell, which is orthogonally oriented to the input polarization to the vapor cell, can be passed and the net rotation is quantified. Similarly, the effect allows for laser locking by the individual analysis of the circular polarization components of the beam. This is known as a dichroic atomic vapor laser lock (DAVLL, Ref.~\cite{cheron_jdpi_1994,corwin_ao_1998}). Here, the circular components are analyzed behind the cell by the combination of a quarter-waveplate and a polarizing beam splitter. Each of the two circular components shows a different spectral shift and the difference of both signals forms dispersive lines for each transition. The zero crossing is usually a reliable lock-point and is used for laser locking~\cite{widmann_ptl_2018}.

%% Slow light
Naturally, also the group velocity of the light is affected by the dispersion in the atomic medium. In the case of a monochromatic input field, it can be determined by calculating the effective refractive index of the different atomic transitions and and calculate the group velocity based on these. Under broadband, i.e.\ non-monochromatic, illumination different spectral components are affected differently. The resulting delay is represented by an integration of the individual delays over the entire spectrum and is generally more complex than in the monochromatic case~\cite{shakhmuratov_pra_2008}. When this is mathematically estimated it becomes clear that a) a light pulse is delayed and smears out more the longer the delay is, and b) that characteristic fingerprints can be observed and the pulse shape is affected in a non-trivial way~\cite{vural_o_2018}.

%% Magnetic field and slow light
The above description of slow light is not limited on a single frequency light input. In addition, when a magnetic field is applied to the hot atomic vapor, both circular components (even when the fields are of the same frequency) are influenced differently. Therefore, in the following both polarization components are analyzed. 

%%%%%%%%%%%% Experimental %%%%%%%%%%%%%%%%%%%%%
%% The experiment
\section{Experiment}

The experimental configuration is shown in Figure~\ref{fig:fig01}. The input light is linearly polarized and the light is analyzed for its circular components with the aid of a quarter wave plate and a polarizing beam splitter. For the initial alignment a laser and commercially photo diodes with variable gain are used, while the atomic cell is at a low temperature, or the laser is several tens of GHz spectrally detuned from the atomic resonance. The light is supplied to the experiment with a single mode fiber. 

%% hot vapor, cell, heater
The atomic vapor cell for these experiments is made of borosilicate glass and has a length of 250~mm. The cell is heated by four round copper blocks which are approx.\ equally spaced along the length of the cell. The two most outer copper blocks heat the cell windows and prevent condensation of the atomic cesium on them. The coldest spot of the cell was aligned with the filling stem of the cell by a piece of aluminum foil which touches the colder parts of the coil from the inside. 

%% Description of Doppler spectrum
When the atomic vapor is heated, an atomic transmission spectrum with Doppler broadened lines is observed. The cesium D$_1$-line shows the well-characterized ground state splitting of 9.192~GHz, plus the excited state splitting of 1.2~GHz. Since the latter is larger than the Doppler broadening of the vapor --at least under ambient conditions-- usually four lines are observed. At higher temperatures, the excited state transitions merge and only two dominant absorption features are observed. Between them, the transmission window shows the typical $1/\delta\nu^2$ detuning frequency dependence and a small window is kept open, which is used to perform the experiments below. This is also the window where slow light can be efficiently observed, since there the effective group index $d n/d \omega$ is approx.\ twice as large as besides the atomic transitions~\cite{camacho_pra_2006}.

%% Magnetic field
It is possible to apply a magnetic field to the cell, such that the Zeeman components are split. This is realized with a long solenoid of enameled copper wire (0.8~mm $\varnothing$). The solenoid is thermally isolated with Teflon supports from the cell heater. After some hours of heating a stable temperature is reached. It is worth mentioning that also the current through the coil also heats the system, which in turn affects the temperature during the application of a magnetic field. This fact becomes relevant when the magnetic field is changed.

%% Quantum dots
The single photon source used here is a strain-tunable In(Ga)As/GaAs quantum dot which is grown by metal-organic vapor-phase epitaxy~\cite{portalupi_nc_2016}. The light-extraction is facilitated by two distributed Bragg reflector (DBR) layers. A pulsed laser with a repetition rate of approx.\ 15~MHz (by pulse picking a standard 80~MHz Ti:Sapphire laser) excites the quantum dot resonantly, while a weak, non-resonant second laser stabilizes the transition. Simultaneously, it prepares a charged exciton transition by exciting charge carriers which initially charge the QD. For single photon detection two standard single photon counting modules were utilized (Excelitas SPCM-AQRH), in combination with time-tagging electronics (Swabian Instruments ``Time Tagger 20'') to evaluate the photon statistics.

%%%%%%%%%%%%%%%%% Results %%%%%%%%%%%%%%%%%%%%%%%
%% Spectra, Temperature determination (Figure 2) 

\begin{figure*}[htb]
  \includegraphics[width=\textwidth]{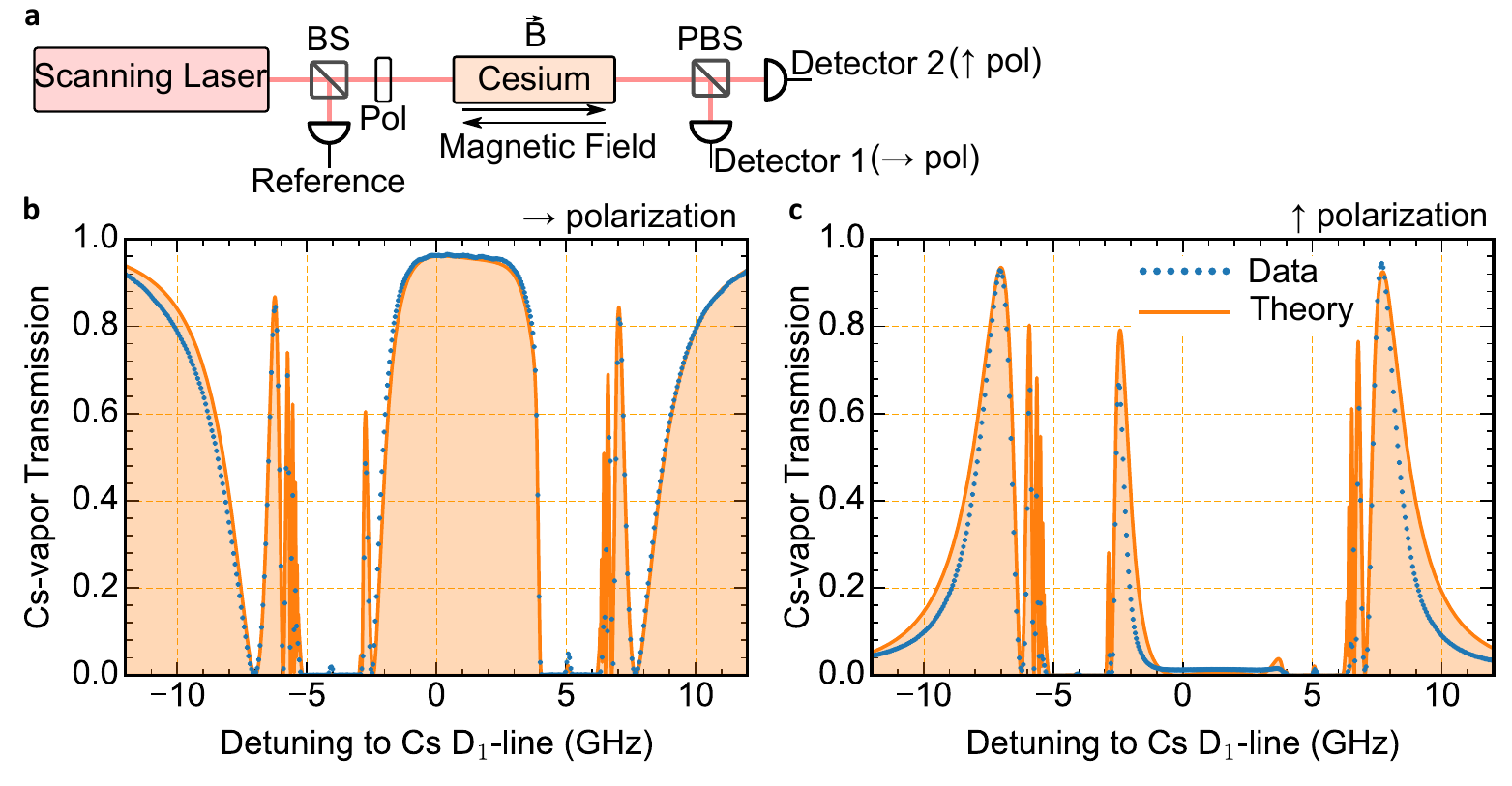}
  \caption{(a) Setup for measuring the Cs-D$_1$ absorption spectra of the polarization components. The polarizer (Pol) ensures that the light enters the vapor cell linearly polarized. The polarizing beamsplitter (PBS) separates the orthogonal polarization components at the two detectors. (b) and (c) Measured (dotted lines) and calculated (solid lines) spectra for the Cs-D$_1$ absorption in a 250~mm long vapor cell at temperature of 80\degree C and a longitudinal magnetic field of 8~mT. The two panels show the separated polarization components: (b) horizontal and (c) vertical. }
  \label{fig:fig02}
\end{figure*}

%% Fig 2. explained
First, we investigate the vapor spectroscopically with an applied magnetic field. Figure~\ref{fig:fig02}a shows the setup where the light enters the vapor horizontally polarized. The light polarization is altered due to the Faraday effect and is detected by a pair of photodiodes behind a polarizing beam splitter. The resulting polarization dependent transmission is exemplarily shown for T=80\degree C and B=8~mT in Figure~\ref{fig:fig02}b. The spectra of both components show oscillating modulations besides the well know absorption profile. The transmission oscillates between zero and the maximal transmission, dictated by the overall vapor absorption. The observed frequency dependent polarization rotation is a consequence of a phase difference between the circular polarization components of the light, due to the circular birefringence induced by the magnetic field. The same effect causes a polarization dependent delay in measurements with pulsed light as observed with laser light on Ref.~\cite{grischkowsky_pra_1973,siddons_np_2009}.

\begin{figure*}[htb]
  \includegraphics[width=\textwidth]{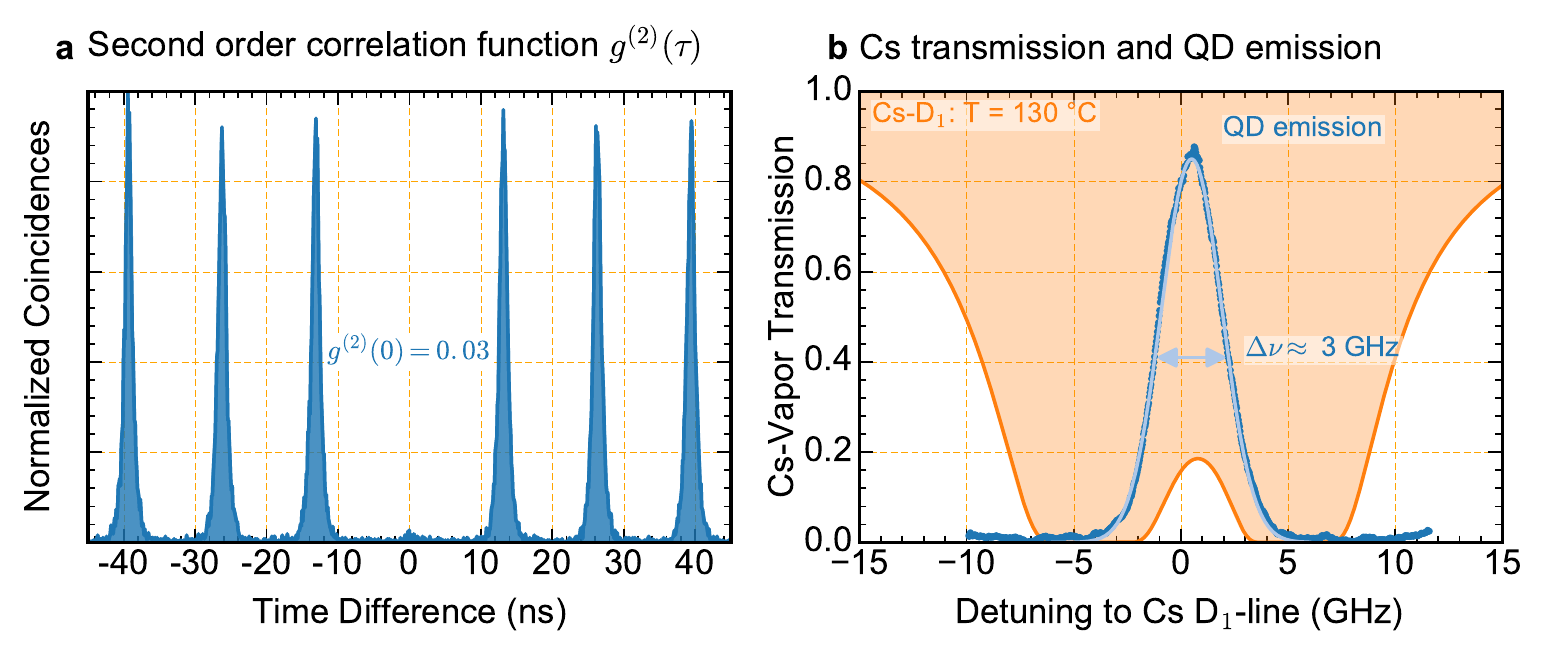}
  \caption{(a) Intensity auto-correlation of the single photons. (b) High resolution resonance fluorescence spectrum of the QD under investigation (blue). For comparison, the calculated absorption spectrum of the Cs-D$_1$ line is shown in the background (orange).}
  \label{fig:fig03}
\end{figure*}

%% Antibunching of quantum dots (Figure 3)
Here, we perform an experiment with single photons under resonant pulsed excitation. Figure~\ref{fig:fig03}a shows the measured second order correlation function of the emission. The vanishing central peak ($g^{(2)}(0)$=0.03) clearly proves the single-photon nature. Figure~\ref{fig:fig03}b shows the spectrum of the QD. A nearly Gaussian profile with 3~GHz width is observed. This well-known shape can be attributed to the presence of spectral diffusion. As comparison, the Cs absorption at 130\degree C is depicted. The width of the QD emission is on the same scale as the width of the transmission window. 

%% Delay (Figure 4)
Due to dispersion in the vapor, the group velocity of photons inside the medium is reduced. Therefore, slow light is observed. In absence of a magnetic field, the delay through the heated atomic vapor amounts to approx.\ 25~ns. We like to note that this delay was similarly observed in our previous work~\cite{vural_o_2018}.

\begin{figure*}[htb]
  \includegraphics[width=\textwidth]{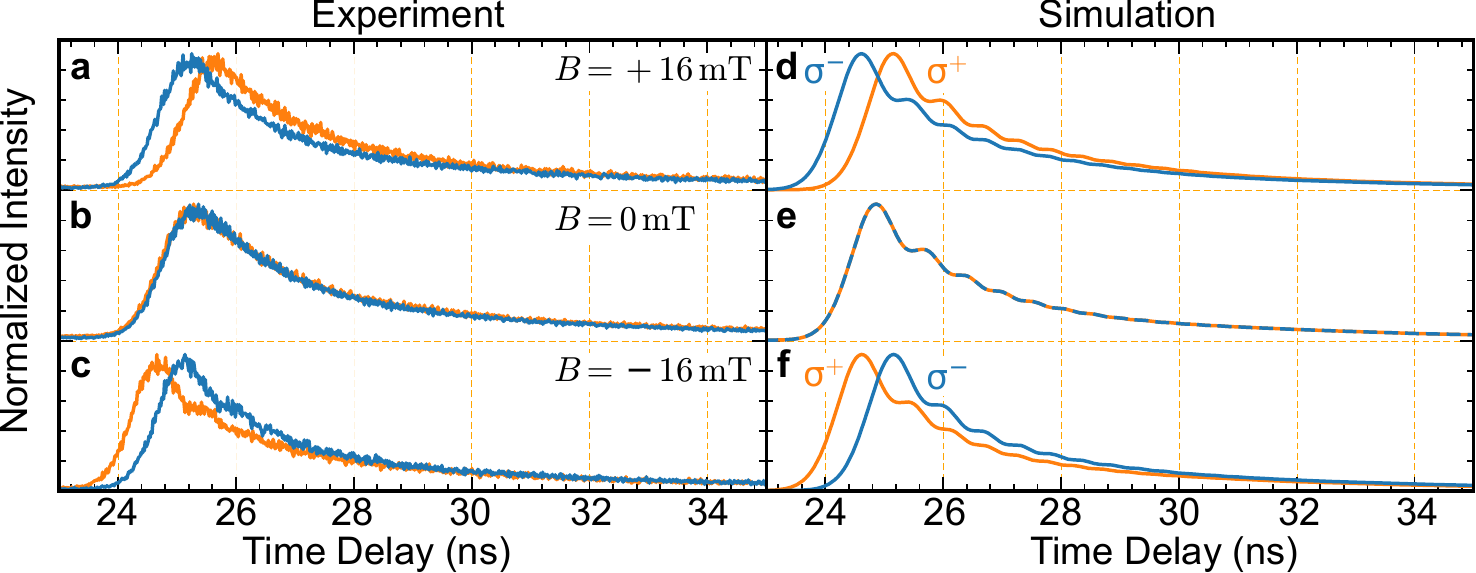}
  \caption{TCSPC measurements with the setup in Figure~\ref{fig:fig01}. Three configurations of the magnetic field: parallel to the direction of propagation (a), without magnetic field (b) and anti-parallel (c). (d)-(f) show the corresponding simulation results. }
  \label{fig:fig04}
\end{figure*}

When a magnetic field is applied, it influences the two circular components of the propagating light differently. This results in different refractive indices and consequently in different group velocities for both components. The setup for the TCSPC measurements is shown in Figure~\ref{fig:fig01}. The QD is excited by resonant $\pi$-pulse. A polarizer before the vapor cell ensures that the photons enter horizontally polarized. Afterwards a $\lambda$/4-plate and a polarizing beam splitter project the both circular components ($\sigma$+ and $\sigma$-) onto two separate APDs. The recorded signals of slow light under different magnetic fields are shown in Figure~\ref{fig:fig04}a-c. While both signals overlap for zero magnetic field (b), one clearly observes the altered delay if a magnetic field of $B$=+16~mT is applied (a). Reversing the orientation of the magnetic field (i.e.\ $B$=-16~mT is applied (c)), simultaneously inverses the delay of the two polarization components. Due to the complex absorption spectrum, the lines are differently affected. The modulations on top of the photon wave packages are due to the vapor's dispersion~\cite{vural_o_2018}. In the experimental signals they occur blurred. The reason are thermal fluctuations which are induced by the coil around the vapor cell. The current through the windings produces additional heat which affects the vapor temperature when the maximum current of $\pm$1~A is applied. On the other hand, the overall acquired delay is almost insensitive to these temperature fluctuations.

%% How the delay calculations were done...
The measurements in Figure~\ref{fig:fig04} are accompanied with theoretical calculations of the delay. The calculations assume a random-walk model of the spectral position of the quantum dot~\cite{vural_o_2018}. Due to fluctuations of the electric and magnetic field in the environment of a QD its energy levels shift randomly -- an effect which is often described as spectral diffusion. Therefore, the carrier frequency of photons changes over time between different emission events. This results in a Gaussian frequency spectrum (see Figure~\ref{fig:fig03}~b) whereas in the Fourier limit a Lorentzian shape is expected. In the simulation model each single photon is assumed to be Fourier limited. That implies an exponential decaying temporal form and a Lorentzian frequency spectrum. The integral of all single shapes with a respective statistical weight results in the profiles of the photon ensemble. The implementation of the conducted simulations follows this picture. The starting point for the propagation of one single photon is the basic Fourier pair of an exponential decay in time domain linked with a Lorentzian shape in frequency domain. Aim of the simulation is to obtain the respective shapes after propagation through the atomic medium. The vapor is regarded as linear medium which allows a full description of the propagation via the complex refractive index. In this case the software tool ElecSus~\cite{Zentile_CPC_2015} was utilized to provide the values of real and imaginary part ($n(\nu)$ and $\alpha (\nu)$). With that it is possible to calculate the spectrum after propagation through the vapor of length $L$
\begin{eqnarray}
\chi_\textnormal{in} (\nu) \rightarrow \chi_\textnormal{out} (\nu) &=& \chi_\textnormal{in} (\nu)\cdot e^{i n_c k L}\\
\mbox{with} \qquad n_c &=& n(\nu) + \frac{i}{2k}\alpha (\nu)\quad .
\end{eqnarray}

\noindent
An inverse Fourier transformation then provides the temporal form of the propagated photons. This procedure is repeatedly performed for an ensemble of photons where each one is assumed to be Fourier limited with a certain carrier frequency. This carrier frequency is drawn from a random Gaussian distribution which is chosen according to the measurement result of the high-resolution spectrum (Figure~\ref{fig:fig03}~b).

%% where the jiggles come from
In the theoretical curves, oscillations superimposed to the exponentially decaying profile are observed. This was already observed experimentally on a comparable configuration~\cite{vural_o_2018} and explained as due to the frequency jitter of the emitted photons (i.e.\ spectral wandering) which propagate through the medium.

\begin{figure*}[htb]
	\centering
	\includegraphics[]{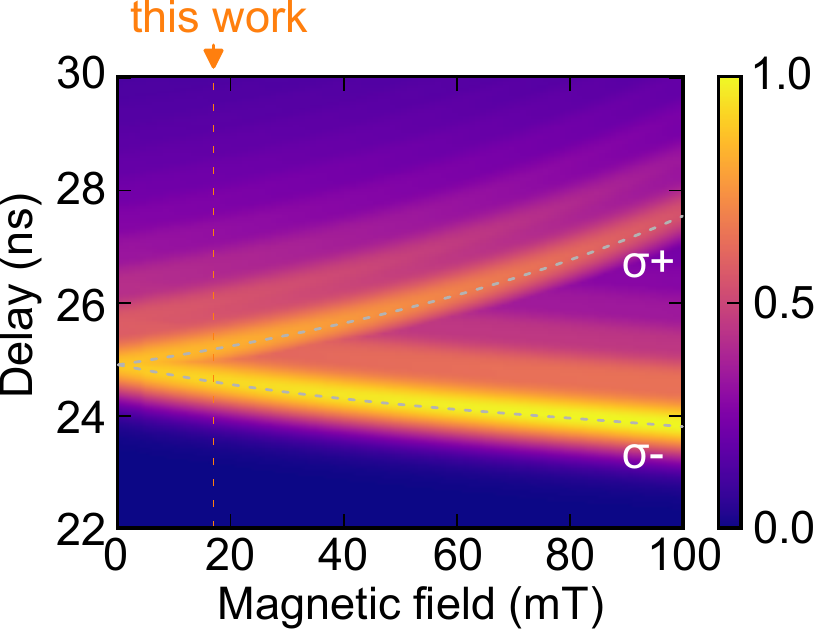}
  \caption{Simulated delay of photons with $\sigma^{+}$- and $\sigma^{-}$-polarization versus the applied magnetic field. The temperature $T=$130\degree C corresponds with the experimental conditions in Figure~\ref{fig:fig04}. The dashed line indicates the operating point in the presented experiment.}
  \label{fig:fig05}
\end{figure*}

%% Different magnetic fields (Figure 5)
The theoretical analysis can be extended, such that larger magnetic fields are covered. This is shown in Figure~\ref{fig:fig05} for the vapor temperature T=130\degree C which is equivalent to the one in the experiment. For zero magnetic fields, the delayed photons have the same arrival time at the single photon detectors. When the magnetic field is increased, the delay is affected and both circular components will not arrive simultaneously, since they are affected by a different refractive index and dispersion. This allows to re-route the photons depending on an external magnetic field.

%% Frequency locking, relation to DAVLL

\section{Conclusion \& Outlook}
%% Conclusion, Outlook %%%%%%%%%%%%%%%%%%%%%%%%%
In conclusion, we have shown experiments with single photons which are delayed and routed by an atomic vapor. The single photons are distributed among the circular analyzed output ports in a time analyzed and modal fashion. This is caused by the reduced group velocity, which behaves differently for both circular components of light. The experiments open the route to fine-tune the delay of single photons from quantum dots with the aid of an atomic cesium vapor.

%% Single Photon Routing, spectral analysis
The slow-light Faraday effect extends over a large spectral region besides the atomic resonances~\cite{siddons_np_2009}. The observed results are intrinsically wavelength-dependent. Therefore, this work provides an additional degree of freedom usable in the comprehension of the emitter dynamics. Therefore, a spectral detuning of the utilized quantum dot, e.g.\ by strain tuning could result in a different relative delay of the quantum dot photons. With this method in turn one can characterize the spectral wandering of the quantum dot emission by a temporal measurement -- a tool, which exceeds the capabilities e.g.\ the slow frequency analysis by a scanning Fabry-P\'erot-cavity.

%% Frequency locking
The shown results of the polarization and frequency selective photon routing is closely related to laser frequency locking. As it has been shown before the introduced technique with the detection of two different circular polarization components of the quantum dot's emission allows for locking a laser to the optical transition~\cite{widmann_ptl_2018}. In this scenario no delay would be required, but the different spectrum of the two polarization components.

%% Anti correlation among two beams
% We like to note that both beams are stricly anti-correlated and will 

\section*{Acknowledgment}
We acknowledge the work of H.\ Kammerlander from the Max Planck Institute's glass workshop who produced the cesium vapor cell. Further the work of W.\ Braun is acknowledged who produced the solenoids and the temperature controller for the cells. We acknowledge funding from the Deutsche Forschungsgemeinschaft in the project GE2737/5-1 and MI300/30-1. S.L.P.\ acknowledges the support of the Baden-W\"urttemberg Stiftung Post-doc Eliteprogramm via the project ``Hybride Quantensysteme f\"ur Quantensensorik''.

\bibliographystyle{unsrt}
% \bibliography{slowlight_01}

\begin{thebibliography}{10}

\bibitem{townsend_el_1993}
P.~D. Townsend, J.~G. Rarity, and P.~R. Tapster.
\newblock Single photon interference in 10 km long optical fibre
  interferometer.
\newblock {\em Electronics Letters}, 29(7):634--635, April 1993.

\bibitem{kimble_prl_1977}
H.~J. Kimble, M.~Dagenais, and L.~Mandel.
\newblock Photon antibunching in resonance fluorescence.
\newblock {\em Phys. Rev. Lett.}, 39:691--695, Sep 1977.

\bibitem{basche_prl_1992}
Th. Basch\'e, W.~E. Moerner, M.~Orrit, and H.~Talon.
\newblock Photon antibunching in the fluorescence of a single dye molecule
  trapped in a solid.
\newblock {\em Phys. Rev. Lett.}, 69:1516--1519, Sep 1992.

\bibitem{gruber_s_1997}
A.~Gruber, A.~Dr{\"a}benstedt, C.~Tietz, L.~Fleury, J.~Wrachtrup, and C.~von
  Borczyskowski.
\newblock Scanning confocal optical microscopy and magnetic resonance on single
  defect centers.
\newblock {\em Science}, 276(5321):2012--2014, 1997.

\bibitem{michler_s_2000}
P.~Michler, A.~Kiraz, C.~Becher, W.~V. Schoenfeld, P.~M. Petroff, Lidong Zhang,
  E.~Hu, and A.~Imamoglu.
\newblock A quantum dot single-photon turnstile device.
\newblock {\em Science}, 290(5500):2282--2285, 2000.

\bibitem{akopian_prl_2006}
N.~Akopian, N.~H. Lindner, E.~Poem, Y.~Berlatzky, J.~Avron, D.~Gershoni, B.~D.
  Gerardot, and P.~M. Petroff.
\newblock Entangled photon pairs from semiconductor quantum dots.
\newblock {\em Phys. Rev. Lett.}, 96:130501, Apr 2006.

\bibitem{young_njp_2006}
Robert~J Young, R~Mark Stevenson, Paola Atkinson, Ken Cooper, David~A Ritchie,
  and Andrew~J Shields.
\newblock Improved fidelity of triggered entangled photons from single quantum
  dots.
\newblock {\em New Journal of Physics}, 8(2):29--29, feb 2006.

\bibitem{unsleber_oe_2015}
Sebastian Unsleber, Christian Schneider, Sebastian Maier, Yu-Ming He, Stefan
  Gerhardt, Chao-Yang Lu, Jian-Wei Pan, Martin Kamp, and Sven H\"{o}fling.
\newblock Deterministic generation of bright single resonance fluorescence
  photons from a purcell-enhanced quantum dot-micropillar system.
\newblock {\em Opt. Express}, 23(26):32977--32985, Dec 2015.

\bibitem{somaschi_np_2016}
N.~Somaschi, V.~Giesz, L.~De~Santis, J.~C. Loredo, M.~P. Almeida, G.~Hornecker,
  S.~L. Portalupi, T.~Grange, C.~Antón, J.~Demory, C.~Gómez, I.~Sagnes, N.~D.
  Lanzillotti-Kimura, A.~Lemaítre, A.~Auffeves, A.~G. White, L.~Lanco, and
  P.~Senellart.
\newblock Near-optimal single-photon sources in the solid state.
\newblock {\em Nature Photonics}, 10:340--, March 2016.

\bibitem{ding_prl_2016}
Xing Ding, Yu~He, Z.-C. Duan, Niels Gregersen, M.-C. Chen, S.~Unsleber,
  S.~Maier, Christian Schneider, Martin Kamp, Sven H\"ofling, Chao-Yang Lu, and
  Jian-Wei Pan.
\newblock On-demand single photons with high extraction efficiency and
  near-unity indistinguishability from a resonantly driven quantum dot in a
  micropillar.
\newblock {\em Phys. Rev. Lett.}, 116:020401, Jan 2016.

\bibitem{michler_book_2017}
Peter Michler, editor.
\newblock {\em Quantum Dots for Quantum Information Technologies}.
\newblock Springer International Publishing, 2017.

\bibitem{phillips_prl_2001}
D.~F. Phillips, A.~Fleischhauer, A.~Mair, R.~L. Walsworth, and M.~D. Lukin.
\newblock Storage of light in atomic vapor.
\newblock {\em Phys. Rev. Lett.}, 86:783--786, Jan 2001.

\bibitem{carruthers_joap_1969}
J.~A. Carruthers and T.~Bieber.
\newblock Pulse velocity in a self-locked hene laser.
\newblock {\em Journal of Applied Physics}, 40(1):426--428, 1969.

\bibitem{grischkowsky_pra_1973}
D.~Grischkowsky.
\newblock Adiabatic following and slow optical pulse propagation in rubidium
  vapor.
\newblock {\em Phys. Rev. A}, 7:2096--2102, Jun 1973.

\bibitem{siddons_np_2009}
Paul Siddons, Nia~C. Bell, Yifei Cai, Charles~S. Adams, and Ifan~G. Hughes.
\newblock A gigahertz-bandwidth atomic probe based on the slow-light {F}araday
  effect.
\newblock {\em Nature Photonics}, 3:225--, March 2009.

\bibitem{siddons_pra_2010}
Paul Siddons, Charles~S. Adams, and Ifan~G. Hughes.
\newblock Optical control of {F}araday rotation in hot {R}b vapor.
\newblock {\em Phys. Rev. A}, 81:043838, Apr 2010.

\bibitem{akopian_np_2011}
N.~Akopian, L.~Wang, A.~Rastelli, O.~G. Schmidt, and V.~Zwiller.
\newblock Hybrid semiconductor-atomic interface: slowing down single photons
  from a quantum dot.
\newblock {\em Nature Photonics}, 5:230--, February 2011.

\bibitem{siyushev_n_2014}
Petr Siyushev, Guilherme Stein, J\"{o}rg Wrachtrup, and Ilja Gerhardt.
\newblock Molecular photons interfaced with alkali atoms.
\newblock {\em Nature}, 509:66--, April 2014.

\bibitem{wildmann_prb_2015}
Johannes~S. Wildmann, Rinaldo Trotta, Javier Mart\'{\i}n-S\'anchez, Eugenio
  Zallo, Mark O'Steen, Oliver~G. Schmidt, and Armando Rastelli.
\newblock Atomic clouds as spectrally selective and tunable delay lines for
  single photons from quantum dots.
\newblock {\em Phys. Rev. B}, 92:235306, Dec 2015.

\bibitem{kiefer_apb_2016}
Wilhelm Kiefer, Mohammad Rezai, J{\"o}rg Wrachtrup, and Ilja Gerhardt.
\newblock An atomic spectrum recorded with a single-molecule light source.
\newblock {\em Applied Physics B}, 122(2):38, Feb 2016.

\bibitem{vural_o_2018}
H\"{u}seyin Vural, Simone~L. Portalupi, Julian Maisch, Simon Kern, Jonas~H.
  Weber, Michael Jetter, J\"{o}rg Wrachtrup, Robert L\"{o}w, Ilja Gerhardt, and
  Peter Michler.
\newblock Two-photon interference in an atom-quantum dot hybrid system.
\newblock {\em Optica}, 5(4):367--373, Apr 2018.

\bibitem{kaneda_a_2018}
Fumihiro Kaneda and Paul~G. Kwiat.
\newblock High-efficiency single-photon generation via large-scale active time
  multiplexing.
\newblock {\em Arxiv}, 2018.

\bibitem{chu_prl_1982}
S.~Chu and S.~Wong.
\newblock Linear pulse propagation in an absorbing medium.
\newblock {\em Phys. Rev. Lett.}, 48:738--741, Mar 1982.

\bibitem{kroh_a_2019}
Tim Kroh, Janik Wolters, Andreas Ahlrichs, Andreas~W. Schell, Alexander Thoma,
  Stephan Reitzenstein, Johannes~S. Wildmann, Eugenio Zallo, Rinaldo Trotta,
  Armando Rastelli, Oliver~G. Schmidt, and Oliver Benson.
\newblock Slow and fast light behavior of single photons from a quantum dot
  interacting with the excited state hyperfine structure of the cesium
  {D}1-line.
\newblock {\em Arxiv}, (1904.08321), 2019.

\bibitem{gerhardt_ol_2018}
Ilja Gerhardt.
\newblock How anomalous is my {F}araday filter?
\newblock {\em Opt. Lett.}, 43(21):5295--5298, 2018.

\bibitem{weller_ol_2012}
L.~Weller, K.~S. Kleinbach, M.~A. Zentile, S.~Knappe, I.~G. Hughes, and C.~S.
  Adams.
\newblock Optical isolator using an atomic vapor in the hyperfine
  {P}aschen-{B}ack regime.
\newblock {\em Opt. Lett.}, 37(16):3405--3407, Aug 2012.

\bibitem{cheron_jdpi_1994}
B~Ch{\'e}ron, H~Gilles, J~Hamel, O~Moreau, and H~Sorel.
\newblock Laser frequency stabilization using {Z}eeman effect.
\newblock {\em Journal de Physique III}, 4(2):401--406, 1994.

\bibitem{corwin_ao_1998}
Kristan~L. Corwin, Zheng-Tian Lu, Carter~F. Hand, Ryan~J. Epstein, and Carl~E.
  Wieman.
\newblock Frequency-stabilized diode laser with the zeeman shift in an atomic
  vapor.
\newblock {\em Appl. Opt.}, 37(15):3295--3298, May 1998.

\bibitem{widmann_ptl_2018}
M.~Widmann, S.~L. Portalupi, P.~Michler, J.~Wrachtrup, and I.~Gerhardt.
\newblock Faraday filtering on the {C}s-{D}1-line for quantum hybrid systems.
\newblock {\em Photonic Technology Letters}, pages 1--1, 2018.

\bibitem{shakhmuratov_pra_2008}
R.~N. Shakhmuratov and J.~Odeurs.
\newblock Slow light with a doublet structure: Underlying physical processes
  and basic limitations.
\newblock {\em Phys. Rev. A}, 77:033854, Mar 2008.

\bibitem{camacho_pra_2006}
Ryan~M. Camacho, Michael~V. Pack, and John~C. Howell.
\newblock Low-distortion slow light using two absorption resonances.
\newblock {\em Phys. Rev. A}, 73:063812, Jun 2006.

\bibitem{portalupi_nc_2016}
Simone~Luca Portalupi, Matthias Widmann, Cornelius Nawrath, Michael Jetter,
  Peter Michler, J\"{o}rg Wrachtrup, and Ilja Gerhardt.
\newblock Simultaneous faraday filtering of the mollow triplet sidebands with
  the {C}s-{D}1 clock transition.
\newblock {\em Nature Communications}, 7:13632--, November 2016.

\bibitem{Zentile_CPC_2015}
Mark~A. Zentile, James Keaveney, Lee Weller, Daniel~J. Whiting, Charles~S.
  Adams, and Ifan~G. Hughes.
\newblock {ElecSus}: A program to calculate the electric susceptibility of an
  atomic ensemble.
\newblock {\em Comput. Phys. Commun.}, 189:162--174, apr 2015.

\end{thebibliography}

\end{document}